\documentstyle[epsfig,prl,twocolumn,aps,floats]{revtex} 
\begin{document}
\topmargin = -2.0cm
\overfullrule 0pt
\twocolumn[\hsize\textwidth\columnwidth\hsize\csname
@twocolumnfalse\endcsname
\title{Seasonal Dependence in the Solar Neutrino Flux}
 \hfill{\small hep-ph/9903473}\\
 \hfill{\small FTUV/99-19, IFIC/99-20} \\
\vglue 0.5cm
\author{P. C. de Holanda$^{1,2}$,
C. Pe\~na-Garay$^{1}$, 
M.\ C.\ Gonzalez-Garcia$^{1}$, 
and
J.\ W.\ F.\ Valle$^{1}$
}
\address{\sl $^1$  Instituto de F\'{\i}sica Corpuscular -- C.S.I.C. \\
    Departamento de F\'{\i}sica Te\`orica, Universitat of Val\`encia \\
    46100 Burjassot, Val\`encia, Spain\\
     http://neutrinos.uv.es  \\
     $^2$ Instituto de F\' {\i}sica Gleb Wataghin\\
    Universidade Estadual de Campinas, UNICAMP
    13083-970 -- Campinas, Brazil}
    \maketitle
\vspace{.5cm}
\hfuzz=25pt
\begin{abstract} 
MSW solutions of the solar neutrino problem predict a seasonal
dependence of the zenith angle distribution of the event rates, due to
the non-zero latitude at the Super-Kamiokande site. We calculate this
seasonal dependence and compare it with the expectations in the
no-oscillation case as well as just-so scenario, in the light of the
latest Super-Kamiokande 708-day data. The seasonal dependence can be
sizeable in the large mixing angle MSW solution and would be
correlated with the day-night effect. This may be used to discriminate
between MSW and just-so scenarios and should be taken into account in
refined fits of the data.
\end{abstract}
\vskip2pc]
\newpage

\medskip

The difference in the $\nu_e$ fluxes during the day and the night due
to the regeneration of the $\nu_e$ in the earth matter -- the so-called
day-night effect -- is one of the milestones of the MSW solutions of
the solar neutrino problem (SNP)~\cite{ms86,daynight}.  This effect is
negligible in the just-so picture~\cite{jso}. Conversely, it is
well-known that vacuum oscillations lead to a seasonal effect due to
the fact that the distance between the Earth and the Sun changes in
different seasons of the year leading to different predicted event
rates beyond the simple geometrical factor.  Though recognized in the
early days of the MSW effect \cite{baltz} the seasonal effect has been
neglected in most discussions of the MSW solution to the SNP and has
even been recently claimed to be absent in the MSW
picture~\cite{justsoseasonal,barger}.

Recent Super-Kamiokande data after 708 days \cite{sk1} exhibit an
excess of the number of events during the night \cite{sk2}.  Though
not yet statistically significant this provides some hint in favor of
the possible existence of a day-night effect. On the other hand there
is also some hint for a seasonal variation in these data, especially
for recoil electron energy above 11.5 MeV. While the former would be
an indication in favour of the MSW solution, the latter would favour
the just-so solution.

\begin{figure}
\begin{center}
\mbox{\epsfig{file=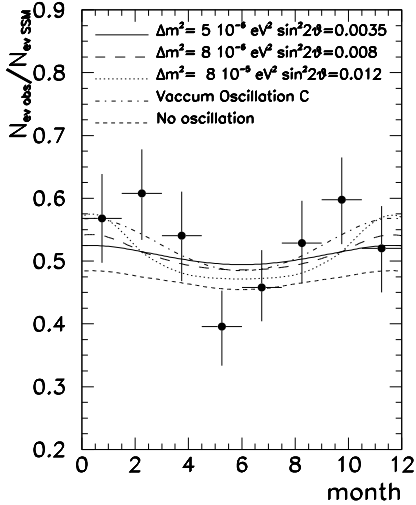,height=7.5cm}}
\end{center}
\vglue -.5cm 
\caption{Ratio of predicted event rate to the SSM prediction 
versus time of the year in Super-Kamiokande for various points in the
SMA solution region of the SNP as labelled.  We have normalized these
three curves to the same yearly averaged event rate which corresponds
to $^8B$ flux normalization 0.7 for the best fit point.  We also show
the expectation in the absence of oscillations with $^8B$ flux
normalization of 0.47 (short dashed line) and the expected effect in
case of vacuum oscillations for solution C in Ref. {\protect
\cite{barger}} (dash-dotted curve) .  The 708 Super-Kamiokande
data points are also displayed.}
\label{sma} 
\end{figure}

\begin{figure}
\begin{center}
\mbox{\epsfig{file=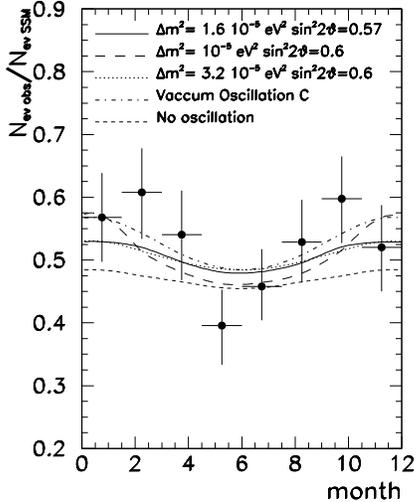,height=7.5cm}}
\end{center} 
\vglue -.5cm
\caption{Ratio of predicted event rate to the SSM prediction 
versus time of the year in Super-Kamiokande for various LMA solutions
of the SNP labelled in the figure.  These three curves are normalized
to the same yearly averaged event rate corresponding to a $^8B$ flux
normalization 1.45 for the best fit point.  We also show the
expectation in the absence of oscillations with $^8B$ flux
normalization of 0.47 (short dashed line) and the expected effect for
vacuum oscillation solution C in Ref. {\protect \cite{barger}}
(dash-dotted curve) together with the 708 Super-Kamiokande data
points.}
\label{lma} 
\end{figure}

Here we call the attention to this interesting feature of the MSW
solution, namely that the expected MSW event rates do exhibit a
seasonal effect due to the different night duration throughout the
year at the experimental site, which leads to a seasonal-dependent $\nu_e$
regeneration effect in the Earth. Taking into account the relative
position of the Super-Kamiokande setup in each period of the year, we
calculate the distribution of the events along the year both for the
large mixing angle (LMA) and the small mixing angle (SMA) solutions to
the SNP.  We find that the effect can be as large as the one expected
in the just-so scenario, especially in the LMA solution, where it
amounts to $\sim 10\%$ at the best fit point for the solar neutrino
event rates given by~\cite{bks}.  For the SMA solution we find that
the magnitude of the seasonal MSW effect is very small at the best fit
point increasing as $\sin^2 2\theta$ increases within the 99\% CL
region.  We illustrate this behaviour in Figs. \ref{sma} and \ref{lma}
and in table \ref{variation}.

Let us now describe our calculation. For simplicity, let us  consider
the two-neutrino mixing case
\begin{equation}
\nu_e = \cos \theta ~\nu_1 + \sin \theta ~\nu_2 \; , \;
\nu_\mu = - \sin \theta ~\nu_1 + \cos \theta ~\nu_2 ~,
\label{eigendef}
\end{equation} 
We have determined the solar neutrino survival probability $P_{ee}$ in
the usual way, assuming that the neutrino state arriving at the Earth
is an incoherent mixture of the $\nu_1$ and $\nu_2$ mass eigenstates.
\begin{equation}
P_{ee} = P_{e1}^{Sun} P_{1e}^{Earth} + P_{e2}^{Sun} P_{2e}^{Earth}
\end{equation}
where $ P_{e1}^{Sun} $ is the probability that a solar neutrino, that
is created as $\nu_e$, leaves the Sun as a mass eigenstate $\nu _1$,
and $ P_{1e}^{Earth} $ is the probability that a neutrino which enters
the Earth as $\nu _1$ arrives at the detector as $\nu_e$. Similar
definitions apply to $P_{e2}^{Sun}$ and $P_{2e}^{Earth}$.

The quantity $P_{e1}^{Sun}$ is given, after discarding the oscillation
terms, as
\begin{equation}
P_{e1}^{Sun}  = 1- P_{e2}^{Sun}  = 
\frac{1}{2} + (\frac{1}{2} - P_{LZ})
cos[2\theta_m(r_0)]
\end{equation}
where $P_{LZ}$ denotes the standard Landau-Zener probability
\cite{LZ} and $\theta_m(r_0)$ is the mixing angle in matter
at the neutrino production point. In our calculations of the expected
event rates we have averaged this probability with respect to the
production point assuming the production point distribution given in
\cite{prod}.

In order to obtain $P_{ie}^{Earth}$ we integrate the evolution
equation in matter assuming a step-function profile of the Earth
matter density.  In the notation of Ref.~\cite{Akhmedov}, we obtain
for $P_{2e}^{Earth}= 1-P_{1e}^{Earth}$
\begin{equation}
P_{2e}^{Earth}(\Phi)=
(Z sin \theta )^2 + (W_1cos \theta + W_3sin \theta )^2\\
\end{equation}
where $\theta$ is the mixing angle in vacuum and the Earth matter
effect is included in the formulas for $Z, W_1$ and $W_3$, which can
be found in Ref.~\cite{Akhmedov}. $P_{2e}^{Earth}$ depends on the
amount of Earth matter travelled by the neutrino in its way to the
detector, or, in other words, on its arrival direction which is
usually parametrized in terms of the nadir angle, $\Phi$, of the sun
at the detector site.

It is very important to realize that the daily range of variation of
the nadir angle depends on the period of the year. As a result the
quantity $P_{2e}^{Earth}$ is seasonal dependent. This will, in turn,
manifest itself as a seasonal dependence of the expected neutrino
event rates.  The general expression of the expected signal in the
presence of oscillations at a given time $t$, $S^{\rm osc}(t)$, is
\begin{eqnarray}
S^{\rm osc}(t)& = & 
\int\! dE_\nu\, \lambda (E_\nu) \times 
\big[ \sigma_e(E_\nu) P_{ee} (E_\nu,t) 
\label{Sosc}
\\ &&+ \sigma_x(E_\nu) 
                        (1-P_{ee} (E_\nu,t))\big] \nonumber,
\end{eqnarray}
where $E_\nu$ is the neutrino energy, $\lambda$ is the neutrino energy
spectrum~\cite{Bspe} with the latest normalization~\cite{SSM},
$\sigma_e$ ($\sigma_x$) is the $\nu_e$ ($\nu_x$, $x=\mu,\,\tau$)
interaction cross section in the Standard Model~\cite{CrSe}, and
$P_{ee} $ is the $\nu_e$ survival probability, which varies in time
through the interval of day and night along the year.  The expected
signal in the absence of oscillations, $S^{\rm no-osc}$, can be
obtained from Eq.(\ref{Sosc}) by substituting $P_{ee}=1$.

The cross sections $\sigma_{e,x}$ are calculated including radiative
corrections and must be corrected for energy threshold and resolution
effects.  In the calculation of the expected signal it is understood
that the $\nu_\alpha$-$e$ cross sections $\sigma_\alpha(E)$ $(\alpha =
e,\,x)$ have to be properly corrected to take into account the
detector energy resolution and the analysis window for each
experiment.  In Super-Kamiokande, the finite energy resolution implies
that the {\em measured\/} kinetic energy $T$ of the scattered electron
is distributed around the {\em true\/} kinetic energy $T'$ according
to a resolution function $Res (T,\,T')$ of the form
\cite{Kr97}:
\begin{equation}
Res (T,\,T') = \frac{1}{\sqrt{2\pi}s}\exp\left[
{-\frac{(T-T')^2}{2 s^2}}\right]\ ,
\end{equation}
where
\begin{equation}
s = s_0\sqrt{T'/{\rm MeV}}\ ,
\label{Delta}
\end{equation}
and $s_0=0.47$ MeV for Super-Kamiokande \cite{sk1,Fo97}. On the other
hand, the distribution of the true kinetic energy $T'$ for an
interacting neutrino of energy $E_\nu$ is dictated by the differential
cross section $d\sigma_\alpha(E_\nu,\,T')/dT'$, that we take from
\cite{CrSe}. The kinematic limits are:
\begin{equation}
0\leq T' \leq {\overline T}'(E_\nu)\ , 
\ \ {\overline T}'(E_\nu)=\frac{E_\nu}{1+m_e/2E_\nu}\ .
\end{equation}

For assigned values of $s_0$, $T_{\rm min}$, and $T_{\rm max}$, the
corrected cross section $\sigma_\alpha(E_\nu)$ is defined as:
\begin{equation}
\sigma_\alpha(E_\nu)=\int_{T_{\rm min}}^{T_{\rm max}}\!dT
\int_0^{{\overline T}'(E_\nu)}
\!dT'\,Res (T,\,T')\,\frac{d\sigma_\alpha(E_\nu,\,T')}{dT'}\ .
\end{equation}

Finally, in order to compare our results with the recent data from
Super-Kamiokande collaboration, we must also include the geometrical
seasonal neutrino flux variation due to the variation of the Sun-Earth
distance ($L \approx 1.5 \times 10^{13}$ cm) arising from the Earth's
orbit eccentricity because the neutrino fluxes in Eq.(\ref{Sosc}) are
yearly averages.  In order to account for this effect we assume a
$1/L^2$ dependence of the flux.  Notice that Super-Kamiokande data are
presented as ratio of observed events over the expected number in the
Standard Solar Model where this expected number of events does not
include the geometrical variation. Thus we must compare the
experimental points with the predictions:
\begin{equation}
\frac{N_{\rm osc}(t_0, \Delta t)}{N_{\rm no-osc}(\Delta t)}=
\frac
{\displaystyle \int_{t_0-\Delta t/2}^{t_0+\Delta t/2} dt\,
\frac{S^{\rm osc}(t)}{\hat{L}^2(t)}
}{\displaystyle \Delta t S^{\rm no-osc}} 
\end{equation}
where
\begin{equation}
\hat{L}(t)=\left[ 1 - \epsilon\cos 2\pi \frac{t}{T}\right]
\end{equation}
and $\epsilon=0.0167$ is the eccentricity of the Earth's orbit around
the Sun, and $T=1$ year.

We now turn to our results. In order to study the behaviour of the
seasonal variation we have explored the parameter space around the
small and large mixing angle solutions, SMA and LMA, respectively. We
find that depending on the values of the mass and mixing angle, one
may get either an enhancement or a damping of the geometrical effect
or even more complicated variations of the signals.

In Fig. \ref{sma} we present the expected event numbers in the recoil
electron energy range from 11.5 MeV up to the maximum, plotted versus
the period of the year for different points in the SMA solution region
of the SNP divided by the BP98 SSM predictions in the absence of
neutrino conversions~\cite{SSM}.  We plot the expected behaviour for
three points: the best fit point obtained by~\cite{bks} with an
arbitrary $^8B$ flux, $\Delta m^2 = 5.\times 10^{-6}$ eV$^2$ and
$\sin^2 2\theta = 3.5 \times 10^{-3}$, a point inside the 99\%
confidence level allowed region with $\Delta m^2 = 8.\times 10^{-6}$
eV$^2$ and $\sin^2 2\theta = 8. \times 10^{-3}$ and a near point with
$\Delta m^2 = 8.\times 10^{-6}$ eV$^2$ and $\sin^2 2\theta = 1.2
\times 10^{-2}$.  We have normalized these three curves to the same
yearly averaged event rate. This corresponds to a $^8B$ flux
normalization 0.7 for the best fit point. For the sake of comparison
we also plot the expected behaviour in the absence of oscillations
with $^8B$ flux normalization of 0.47 as well as the best fit point
for the vacuum solution C of Ref. \cite{barger}.

As seen in the figure the seasonal effect is comparable to the
expectation in the absence of oscillation at the best fit point of the
SMA solution and it increases as the mixing angle increases.
In Table \ref{variation} we show the seasonal variation (in percent)
defined as
\begin{displaymath}
Var \equiv 2 \frac{R_{max}-R_{min}}{R_{max}+R_{min}} 
\end{displaymath}
for the different MSW and vacuum solutions of the SNP where $R(t)=
N_{\rm osc}(t)/N_{\rm SSM}$. We find that for the SMA solution the
effect increases as one increases $\sin^2 2\theta$. For example for
$\sin^2 2\theta = 0.008$, still within the 99 \% CL allowed region, it
reaches 10\% and for $\sin^2 2\theta = 0.012$ it gets to be as large
as 20\%.  Of course, since the seasonal effect is induced by the
variation of the regeneration in the Earth along the year, the effect
is large only in the parameter region where the day-night effect is
not negligible, which corresponds to larger mixing angle values. Note
that in the SMA region the points we have chosen in order to
illustrate the possible seasonal variation in the MSW picture are
consistent with the measured yearly average day-night asymmetry.
\begin{table}
\begin{tabular}{|c|c|c|c|}
Point & $\Delta m^2$ (eV$^2$) &  $\sin^2(2\theta)$ &  Var  (\%)\\
\hline
No-oscillation &  & & 6 \\
\hline
MSW SMA  &  & & \\
\hline
Best Fit Point 
& $5\times 10^{-6}$ & $3.5 \times 10^{-3}$ & 6 \\
& $8\times 10^{-6}$ & $8 \times 10^{-3}$ & 10 \\
& $8\times 10^{-6}$ & $1.2 \times 10^{-2}$ & 20 \\
\hline
MSW LMA  &  & & \\
\hline
Best Fit Point 
& $1.6\times 10^{-5}$ & $0.57$ & 10 \\
& $1.\times 10^{-5}$ & $0.6$ & 22 \\
& $3.2\times 10^{-5}$ & $0.6$ & 9 \\
\hline 
Vacuum Solutions &  & & \\
\hline
 C 
& $4.4\times 10^{-10}$ & $0.93$ & 15 \\
 D
& $6.4\times 10^{-10}$ & $1$ & 12 \\
 A 
& $6.5\times 10^{-11}$ & $0.7$ & 9 \\
\end{tabular}
\vglue 0.4cm
\caption{Seasonal variation (in percent) of the ratio of predicted 
event rate in various oscillation scenarios to the SSM prediction.}
\label{variation}
\end{table}

Now we turn to the LMA solution of the SNP where the effects are
potentially larger. Our results for this case are displayed in
Fig. \ref{lma}.  Again, we plot the expected behaviour for three
characteristic points: the best fit point obtained by~\cite{bks} with
an arbitrary $^8B$ flux ($\sin^2 2\theta = 0.57$, $\Delta m^2 =
1.6\times 10^{-5}$ eV$^2$), a point inside the 99\% confidence level
allowed region with $\Delta m^2 = 1.\times 10^{-5}$ eV$^2$ and $\sin^2
2\theta = 0.6$ and a point inside the allowed region where the
expected average day-night asymmetry is smaller, $\Delta m^2 =
3.2\times 10^{-5}$ eV$^2$ and $\sin^2 2\theta = 0.6$. We have
normalized these three curves to the same yearly averaged event
rate. This corresponds to a $^8B$ flux normalization 1.45 for the best
fit point. We also plot the expected behaviour in the absence of
oscillations with $^8B$ flux normalization of 0.47 and the best fit
vacuum solution C from Ref. \cite{barger}. In Table \ref{variation} we
show the variation (in percent) corresponding to these points.  As seen
in the table the effect at the best fit point of the LMA solution (10
\%) is comparable with the corresponding effect in some of the
favoured vacuum oscillation solutions.  In the LMA solution region the
seasonal variation is very mildly dependent on the mixing angle while
presents an oscillatory variation with $\Delta m^2$. Our results show
that depending on the mass and mixing angle values one may get an
enhancement or damping of the geometrical effect. We must bear in mind,
however, than in the lower $\Delta m^2$ part of the LMA solution
region, the expected yearly average day-night asymmetry is in conflict 
with the existing data.  

Finally let us comment on the effect of an enhanced hep neutrino flux
as suggested in ~\cite{bk98} in order to account for the recent
Super-Kamiokande measurements of the energy spectrum. We find that
even with large hep enhancement factors of 20 or more, the expected
modifications of our results near the best fit points both for the SMA
and LMA are small.

To summarize, we have shown that MSW solutions of the solar neutrino
problem can lead to a sizeable seasonal dependence of the event rates
in the large mixing angle region and this should be taken into account
in refined fits of the data where the day-night analysis is also
performed. The MSW seasonal effect is correlated with the day-night
asymmetry and may potentially be useful in order to pinpoint the
underlying mechanism involved in the explanation of the solar neutrino
anomaly, discriminating between different solutions. For example, the
non-observation of the day-night effect and the confirmation of
seasonal-dependent rates would provide an indication for the just-so
picture. Conversely, a possible confirmation of a seasonal dependence
accompanied by the day-night effect would point towards a LMA MSW-type
solution.

\vskip .4cm

We are grateful to E. Akhmedov and C. Yanagisawa for useful comments.  
This work was
supported by Spanish DGICYT under grant PB95-1077, by the European
Union TMR network ERBFMRXCT960090. P. C. de Holanda was supported by
FAPESP (Brazil).

\end{document}